# Unsaturated Single Atoms on Monolayer Transition Metal Dichalcogenides for Ultrafast Hydrogen Evolution

Yuting Luo[1†], Shuqing Zhang[1†], Haiyang Pan[2], Shujie Xiao[3], Zenglong Guo[4], Lei Tang[1], Usman Khan[1], Baofu Ding[1], Meng Li[5], Zhengyang Cai[1], Yue Zhao[2], Wei Lv[3], Qinliang Feng[5], Xiaolong Zou[1*], Junhao Lin[4*], Hui-Ming Cheng[1, 6], & Bilu Liu[1*]

[1] Shenzhen Geim Graphene Center (SGC), Tsinghua-Berkeley Shenzhen Institute (TBSI), Tsinghua University, Shenzhen 518055, P. R. China.

[2] Institute for Quantum Science and Engineering, Southern University of Science and Technology, Shenzhen 518055, P. R. China.

[3] Graduate School at Shenzhen, Tsinghua University, Shenzhen 518055, P. R. China.

[4] Department of Physics, Southern University of Science and Technology, Shenzhen 518055, P. R. China.

[5] MOE Key Laboratory of Material Physics and Chemistry under Extraordinary Conditions, Shaanxi Key Laboratory of Optical Information Technology, School of Science, Northwestern Polytechnical University, Xi'an 710072, P. R. China.

[6] Shenyang National Laboratory for Materials Science, Institute of Metal Research, Chinese Academy of Sciences, Shenyang 110016, P. R. China.

[†] These authors contribute equally.

Correspondence should be addressed to B.L. (email: bilu.liu@sz.tsinghua.edu.cn), J.L. (email: linjh@sustech.edu.cn), or X.Z. (email: xlzou@sz.tsinghua.edu.cn).

**SUMMARY**

**Large scale implementation of electrochemical water splitting for hydrogen evolution requires cheap and efficient catalysts to replace expensive platinum. Molybdenum disulfide is one of the most promising alternative catalysts but its intrinsic activity is still inferior to platinum. There is therefore a need to explore new active site origins in molybdenum disulfide with ultrafast reaction kinetics and to understand their mechanisms. Here, we report a universal cold hydrogen plasma reduction method for synthesizing different single atoms sitting on two-dimensional monolayers. In case of molybdenum disulfide, we design and identify a new type of active site, *i.e.,* unsaturated Mo single atoms on cogenetic monolayer molybdenum disulfide. The catalyst shows exceptional intrinsic activity with a Tafel slope of 35.1 mV dec$^{-1}$ and a turnover frequency of ~$10^3$ s$^{-1}$ at 100 mV, based on single flake microcell measurements. Theoretical studies indicate that coordinately unsaturated Mo single atoms sitting on molybdenum disulfide increase the bond strength between adsorbed hydrogen atoms and the substrates through hybridization, leading to fast hydrogen adsorption/desorption kinetics and superior hydrogen evolution activity. This work shines fresh light on preparing highly-efficient electrocatalysts for water splitting and other electrochemical processes, as well as provides a general method to synthesize single atoms on two-dimensional monolayers.**

**INTRODUCTION**

Concerns about the energy crisis and environmental pollution have called for the development of hydrogen, a clean and high density energy carrier, as a key alternative to fossil fuels.[1,2] Currently, hydrogen has been mainly produced using carbon-based resources, *i.e.*, natural gas, oil, and coal, making carbon emission unavoidable.[3] The production of hydrogen by electrochemical water splitting promises to overcome this challenge, especially when the reaction is driven by electricity generated by renewable energy such as solar and wind. However, the hydrogen evolution reaction (HER) has a much lower thermal efficiency than its thermodynamic limit, resulting in the need of using highly efficient electrocatalysts. Platinum (Pt) has been proved to be the most efficient HER catalyst, but its scarcity and high cost hamper its wide use. In recent years, various Pt-free materials have been explored as alternative HER electrocatalysts, including metal alloys,[4,5] metal hydroxides,[6,7] metal disulfides,[8-10] metal carbides,[11,12] and metal phosphides[13,14]. Nevertheless, the intrinsic HER activities of these catalysts are still inferior to Pt. Increasing the numbers of active sites may endow catalysts with decent HER performance, but also make the mechanism study difficult. Noteworthy, monolayer two-dimensional (2D) materials possess unambiguous crystal structures and accurately determined reaction area[15-17], *i.e.*, nearly identical interface area and electrode area, serving as ideal models to investigate intrinsic activity of catalysts (**Figure. S1**). Among various 2D materials, molybdenum disulfide ($MoS_2$) is promising due to its earth abundance and good stability.[18] Interest in $MoS_2$ catalyst can be traced back to the 1970s in the hydrodesulphurization (HDS) processes, where it was used to remove sulfur and add hydrogen in hydrocarbon streams.[19,20] As HER shares a similar H* intermediate with the HDS reaction, $MoS_2$ was also found to be active for the HER in 2007, where the edges of 2D $MoS_2$ were found to be the active sizes.[21,22] Later, it was shown that phase transformation of $MoS_2$ from the stable 2H phase to the metastable 1T phase activates the inert basal plane of $MoS_2$ for the HER.[23,24] The sulfur vacancies (S vacancies) and strain in $MoS_2$ were later found to be active for the HER. [25,26] After that, researchers have devoted much effort to synthesize 2D $MoS_2$ for better HER activity by creating more active sites including edges, the 1T phase, S vacancies, or strain (**Figure**

**1A**).[27-31] However, the HER activities of 2D $MoS_2$ (usually with overpotentials larger than 170 mV at 10 mA cm$^{-2}$ and Tafel slopes larger than 40 mV dec$^{-1}$) are still inferior to Pt-based catalysts (**Table S1**).[15,30,32-35] This calls for the search for new active sites in $MoS_2$ to further improve its HER performance.

On the other side, single atoms (SAs) have become important in recent years, because they show great potential as highly efficient catalysts with a higher catalytic activity per atom site and better stability than nanoparticle catalysts.[36] SAs have the simplest structures usually with only one atom as the active site, and thus in principle, could be ideal for studying the mechanism of catalytic process. For example, multinuclear active sites are commonly necessary for water oxidation, while recently Guan *et al.*[37] reported that Mn SAs embedded in nitrogen-doped graphene is effective in the oxygen evolution reaction, suggesting that a mononuclear active site is capable for oxygen evolution. Note that although SAs have well-defined structures, they are usually loaded on bulk or multilayer supports with different structures, making their microscopic chemical environment, coordination conditions, and structures complicated. Ideally, one can use supports with clear and well-defined structures such as 2D monolayers to support SAs. On one hand, this will simplify the system which in turn benefits the mechanism study. This is because monolayer supports are suitable for obtaining better statistical density of SAs, not only the mass density (wt%) and atom density (at%) widely used for SAs on porous or bulk supports in literature, but also the area density (atom nm$^{-2}$) and volume density (atom nm$^{-3}$) which are only accurate in atomically thin 2D supports. On the other hand, for cases like 2D $MoS_2$, combining SAs with monolayer $MoS_2$ may bring opportunities to create new types of active sites to further improve its HER activity. Unfortunately, the synthesis of metal SAs on 2D monolayer supports is still a grand challenge and has rarely been reported.

Here, we develop a cold hydrogen plasma reduction method and successfully synthesize different SAs/2D monolayer systems, where SAs sit on 2D monolayers. Our idea is inspired by the fact that hydrogen plasma shows a strong reducibility and in principle can dissociate chemical bonds between anions and metal cations in metal compounds.[38-40] For an exemplified system of Mo SAs sitting on cogenetic $MoS_2$ monolayers (denoted as Mo SAs/ML-$MoS_2$), experimental and theoretical results show that the Mo SAs are coordinately unsaturated since Mo SAs transfer less charge to sulfur atoms than do saturated Mo atoms in the pristine $MoS_2$. We use this Mo SAs/ML-$MoS_2$ system as a HER model catalyst (**Figure 1B**) based on the following considerations. First, it is assumed that the active sites on $MoS_2$ for HDS are unsaturated Mo centers,[18] showing certain similarity to HER activity given by unsaturated Mo atoms near S vacancies.[24] We therefore expect unsaturated Mo SAs to have a decent HER activity. Second, a Mo SA on monolayer $MoS_2$ has a simple and well-defined structure, which is suitable for the study of its HER kinetics and understanding the mechanism. Electrochemical measurements based on single $MoS_2$ flake microcells show that Mo SAs/ML-$MoS_2$ catalysts show ultrafast HER kinetics with a small Tafel slope of 35.1 mV dce$^{-1}$ and an impressively high turnover frequency (TOF) of ~$10^3$ s$^{-1}$ at 100 mV. By the analyses of the density of states, we find that an unsaturated Mo SA on $MoS_2$ effectively hybridizes with H, leading to increased H bonding and a lower Fermi level for the hydrogen adsorption compared to saturated Mo systems. An almost zero hydrogen adsorption free energy is achieved in Mo SAs/ML-$MoS_2$, giving rise to the superior intrinsic activity for the HER.

**RESULTS**

As shown in **Figure 1C**, we developed a new and universal strategy for treating 2D monolayers by a cold hydrogen plasma to obtain metal SAs on 2D monolayers. In principle, when conditions are controlled properly, the hydrogen plasma which has a stronger reducing ability than hydrogen molecules,[38,39] may bond with anions like sulfur atoms and dissociate metal atoms at a low temperature, generating metal SAs on 2D monolayers. Take Mo SAs/ML-MoS$_2$ system as an example, where Mo SAs sitting on the monolayer MoS$_2$ (**Figure 1B**). The Mo SAs/ML-MoS$_2$ system was synthesized by a two-step method (see details in the "**Methods**" section and **Figure S2**). First, ML-MoS$_2$ was grown on a SiO$_2$/Si substrate by chemical vapor deposition (CVD) at a growth temperature of 750 °C. The monolayer nature of the MoS$_2$ was demonstrated by atomic force microscopy (AFM), photoluminescence (PL), and Raman spectroscopy (**Figure S3**). Second, the sample was treated by the hydrogen plasma under different conditions. Besides MoS$_2$, we prepared W SAs on monolayer WS$_2$ (denoted as W SAs/ML-WS$_2$, **Figure 1D**) by this method (see details in **Supplemental Information**). We studied several parameters in the plasma process including temperature, pressure, time and radio frequency power, and found that temperature especially played a key role. When the temperature is above 200 °C, dissociation of the Mo-S bonds becomes easy and large holes appear in the monolayer MoS$_2$ (**Figure S4**), consistent with a previous report.[41] Therefore, we control the temperature to be below 80 °C, which is monitored by a thermocouple in the plasma chamber, to avoid destruction of the structure of the MoS$_2$ flakes and aggregation of Mo atoms. Under proper conditions, no obvious structural changes like holes or cracks appear in the Mo SAs/ML-MoS$_2$ samples (**Figure S5**). The best sample was prepared with a H$_2$ flow rate of 20 standard cubic centimeters per minute (sccm) and a radio frequency power smaller than 30 W for 80 s, which contained abundant Mo SAs on monolayer MoS$_2$. The key to synthesize Mo SAs/ML-MoS$_2$ is to control the reaction by kinetics rather than thermodynamics. MoS$_2$ flakes prepared by the Scotch tape exfoliation method behaved the same as the CVD-grown MoS$_2$ upon hydrogen plasma treatment (**Figure S6**). We also examined the use of an Ar plasma instead of the hydrogen plasma (under similar pressure, time, power of radio frequency, and temperature), and found that the degree of reduction of the Mo decreases, showing that hydrogen is necessary because of its strong reducing ability being able to chemically dissociate the Mo-S bonds in MoS$_2$ (**Figure S7**).

To study the microscopic structure of the SAs/2D monolayer samples, high-angle annular dark-field STEM (HAADF-STEM) characterization was performed on both pristine MoS$_2$ and Mo SAs/ML-MoS$_2$ samples for comparison. As shown in **Figure S8A**, the pristine MoS$_2$ shows a hexagonal crystal structure with alternating bright and dark spots, which are assigned to columns of Mo and S$_2$ according to their atomic weight. We occasionally observe a few sites whose intensity is higher than that of their neighbors (**Figure S8A**). Combined with energy dispersive spectroscopy (EDS), X-ray photoelectron spectroscopy (XPS), and image simulation (**Figures. S9-S11**), these brighter sites are recognized as Mo SAs. Similar results were also reported by Hong *et al*, suggesting that trace amount of Mo adatoms present in CVD-grown MoS$_2$ samples.[42]. The density of Mo SAs in the pristine MoS$_2$ was measured to be 0.0051 atoms nm$^{-2}$, while after hydrogen plasma treatment the HAADF-STEM images of Mo SAs/ML-MoS$_2$ sample show that the density of Mo SAs has increased to 0.4365 atoms nm$^{-2}$, almost two orders of magnitude higher than that of the pristine MoS$_2$ sample (**Figure 1D**, **Figures. S8B and S8C**). The Mo SAs/ML-MoS$_2$ sample shows SA densities that are similar to recent reports, having a mass density of 2.21 wt% and an atom density of 1.23 at% (**Table S2**). Moreover, thanks to the clear crystal structure of the ML-MoS$_2$ support, we estimated the area density of Mo SAs on Mo SAs/ML-MoS$_2$ to be 4.35 × 10$^{13}$ atoms cm$^{-2}$, which cannot be estimated in the case of porous or bulk supports.

We also found that the Mo SAs/ML-MoS$_2$ samples retain the well-defined hexagonal lattice without noticeable damage to the basal planes of the original lattice. To further verify the integrity of the MoS$_2$ lattice, we compared the densities of S vacancies for pristine MoS$_2$ (0.14 nm$^{-2}$) and Mo SAs/ML-MoS$_2$ (0.24 nm$^{-2}$) samples, which are quite low and close to each other.[26,35,42] Meanwhile, Mo vacancies were seldomly found in the Mo SAs/ML-MoS$_2$ sample which indicates that the Mo SAs may originate from the edges of the MoS$_2$ flakes, due to their lower cohesive energy than those in the basal plane. Similarly, a considerable amount of W SAs (1.76 × 10$^{13}$ atoms cm$^{-2}$) were synthesized on monolayer WS$_2$ after the same plasma treatment (**Figure 1E**). Taken together, these results demonstrate that the metal SAs/2D monolayer systems with a high loading of different metal SAs were synthesized after the cold hydrogen plasma reduction of different 2D monolayers.

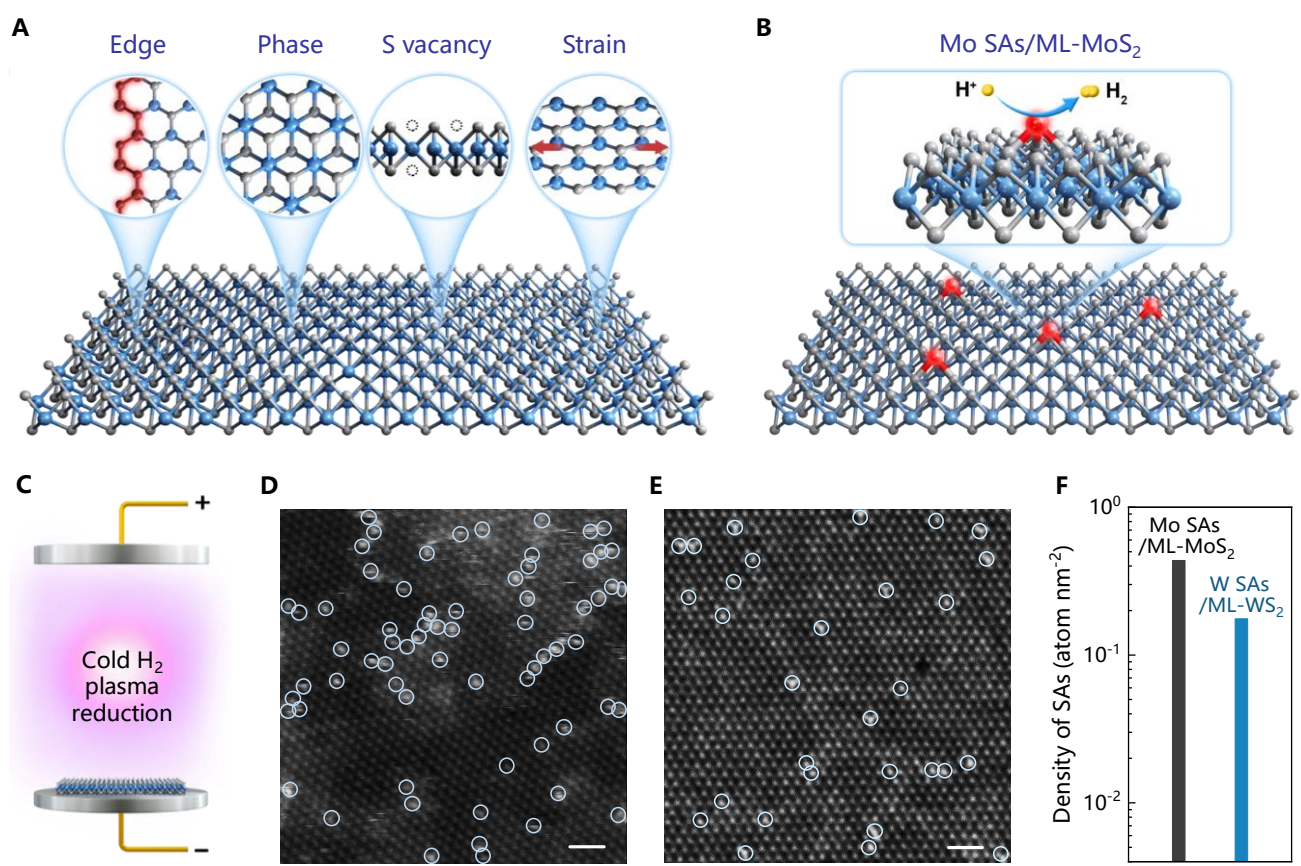

**Figure 1. Formation of metal SAs on 2D monolayers.** (A) Schematic showing representative active

sites of MoS$_2$ for hydrogen evolution reported previously, including edges, S vacancies, 1T phase, and strain. (B) Schematic showing Mo SAs/ML-MoS$_2$ as a new active site for hydrogen evolution. (C) Schematic showing the synthesis of metal SAs on cogenetic 2D monolayer by cold hydrogen plasma reduction. (D,E) HAADF-STEM images of (D) Mo SAs/ML-MoS$_2$ and (E) W SAs/ML-WS$_2$ samples, showing high densities of Mo and W SAs in the samples. The scale bar is 1 nm for d and e. (F) Statistical results of the density of Mo and W SAs on 2D monolayers.

To further confirm that the bright spots are Mo single atoms on ML-MoS$_2$ and to study their local configuration, we carried out detailed HAADF-STEM analyses. The full width at half maximum (FWHM) of the intensity profile of each brighter site was collected from the HADDF-STEM images, and statistical analysis showed a narrow size distribution with an average size of 0.16 nm for single atoms resolved in 60 kV STEM images (**Figure 2A**). This average size is slightly larger than that of Mo (IV) atoms (0.14 nm) in the MoS$_2$ basal plane, which may be attributed to the larger Debye–Waller factor of a Mo SA due to its unsaturated bonding. Intensity profile analysis of an enlarged HAADF–STEM image revealed (**Figure 2B)** that the brighter sites (blue spot marked with 2) have nearly double the intensity of the Mo sites in the basal plane of MoS$_2$ (blue spot marked with 1), suggesting a Mo SA sitting on top of a Mo atom of MoS$_2$ (**Figure 2B**). Moreover, the corresponding fast Fourier transform (FFT) pattern (**Figure 2C**) shows no obvious changes of the lattice parameters along the [0001] zone axis, suggesting that the Mo SAs have a negligible effect on the lattice structure of underlying monolayer MoS$_2$. Careful examination on tens of Mo SAs showed that the Mo SAs/ML-MoS$_2$ catalyst contained two types of Mo SAs, one sitting on Mo atop sites (Mo SA-1 in **Figure 2D**) and the other sitting above S vacancy sites (Mo SA-2 in **Figure S12**) in the MoS$_2$ basal plane. Statistical analysis shows that around two thirds of the Mo SAs are type 1 and one third are type 2. The structures of these two types were studied by density functional theory (DFT) calculations to understand their coordination status, and their relaxation structures and a comparison with experimental results are shown in **Figure S13**. As discussed later, both configurations were found to be active in the HER reaction. Similarly, we also carried out detailed HAADF-STEM analyses and simulations on the W SAs/ML-WS$_2$ samples, showing that W SAs are sitting either on the atop site of W atoms in the WS$_2$ or above S vacancy sites (**Figures S14 and S15**).

Taking Mo SAs/ML-MoS$_2$ system as an example, we studied the samples and found that unlike saturated Mo atoms in MoS$_2$ basal plane where each Mo atom bond with six sulfur atoms, the Mo SAs on the basal plane of MoS$_2$ are unsaturated. To study their chemical environment in the Mo SAs/ML-MoS$_2$ sample, we carried out high resolution XPS measurements. **Figure 2E** shows that the Mo 3$d$ peak of the Mo SAs/ML-MoS$_2$ sample shifts greatly to a lower binding energy than that of the pristine MoS$_2$ sample. For example, a large binding energy shift of the Mo 3$d$ $_{2/5}$ peak (-0.57 eV) is observed for a sample after hydrogen plasma treatment for 80 s (**Figure S16**). In addition, the Mo 3$d$ $_{2/5}$ and Mo 3$d$ $_{2/3}$ peaks show a larger FWHM because of the existence of both Mo (IV) in the MoS$_2$ basal planes and Mo SAs on the surface with a smaller binding energy (**Figure 2E**). Each Mo (IV) atom in the basal plane is bonded to six S atoms, while a Mo SA only bond to three S atoms or bond to two S atoms and leaves a S vacancy. As a result, the S/Mo ratio decreases with increasing plasma treatment time, as revealed by both XPS and STEM results (**Figure S17**). To summary the unsaturated coordination of Mo single atoms on monolayer MoS$_2$, **Figure 2F** shows the ratio of the number of unsaturated Mo atoms in the Mo SAs/ML-MoS$_2$ and pristine MoS$_2$ samples to the total number of Mo atoms, where unsaturated Mo atoms are defined as those that are not bonded to six S atoms (*i.e.*, Mo SAs and Mo atoms in the basal plane of $MoS_2$ adjacent to S vacancy). Moreover, the calculated Bader charges (**Figure 2G**) show that the all Mo atoms are positively charged. A Mo atom in pristine $MoS_2$ loses 1.06 *e*, which is more than the 0.83 *e* for Mo SA-1 and 0.58 *e* for Mo SA-2. Therefore, Mo SAs transfer less charge to sulfur atoms than do saturated Mo atoms in pristine $MoS_2$, showing that the Mo SAs are unsaturated. Both experiment and theoretical calculations give consistent results, that is, unsaturated Mo SAs are formed on the $MoS_2$ in Mo SAs/ML-$MoS_2$ samples.

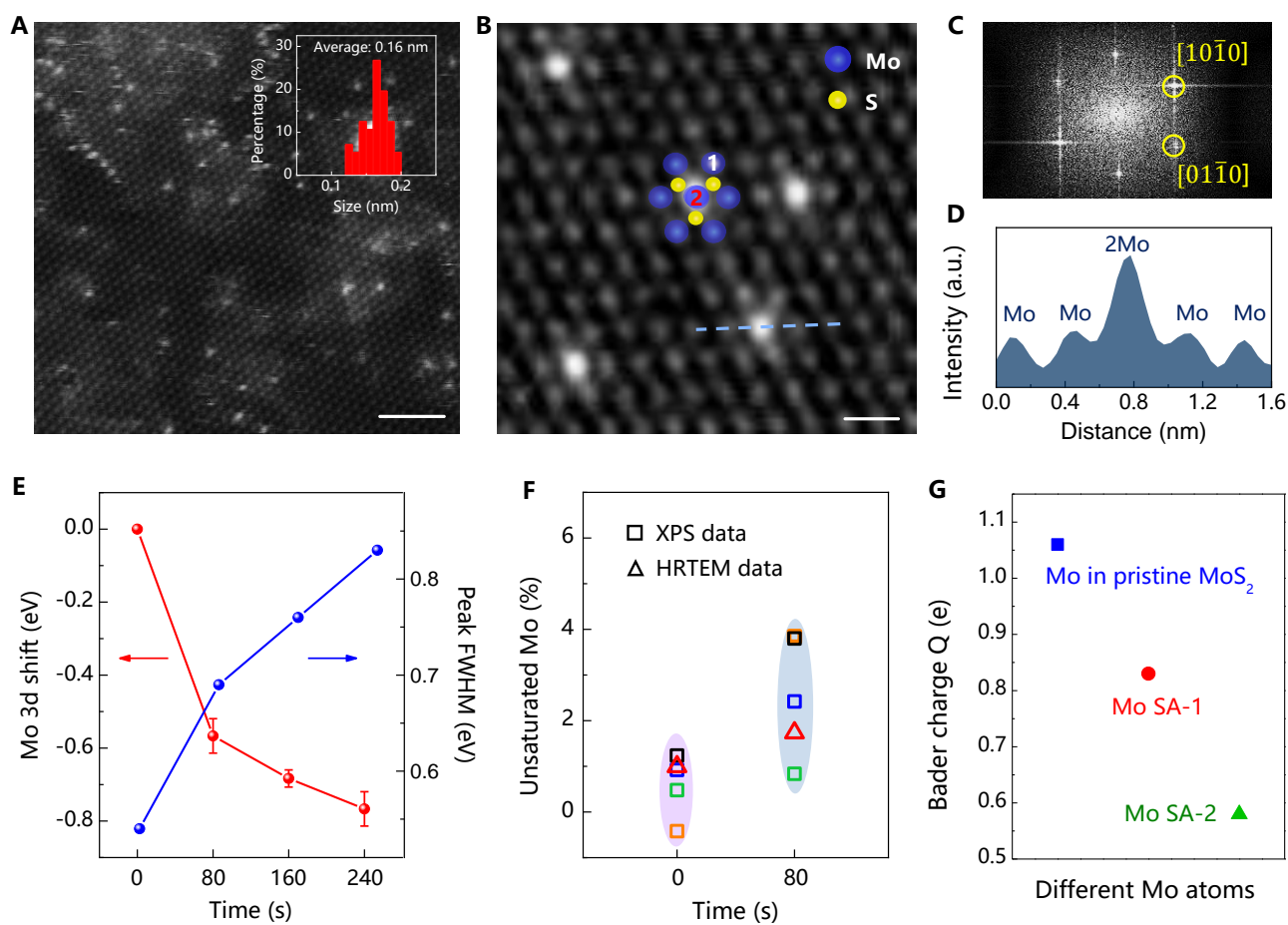

**Figure 2. Structural and chemical characterization of the Mo SAs/ML-MoS₂ samples.** (A) HAADF-STEM image and size distribution of the Mo SAs. (B) Enlarged HAADF-STEM image, (C) corresponding FFT pattern, and (D) intensity profile along the blue line in B. The results indicate that Mo SAs sitting on top of Mo atoms in the MoS$_2$ basal planes. (E) Statistical results of the positions and FWHM of the Mo 3d$_{3/2}$ XPS peak for MoS$_2$ samples treated in hydrogen plasma for different time. Data were measured three times for each sample, and error bars correspond to the standard deviations. (F) XPS and HRTEM results showing the ratio of unsaturated Mo to all Mo atoms for Mo SAs/ML-MoS$_2$ samples. The XPS data was collected from four different samples for both pristine MoS$_2$ and Mo SAs/ML-MoS$_2$. (G) Bader charges of Mo atoms in different configurations, including Mo in the pristine MoS$_2$ (blue square), Mo SA sitting on the Mo atop sites of MoS$_2$ basal plane (*i.e.*, Mo SA-1, red circle), and Mo SA sitting above the S vacancy sites (*i.e.*, Mo SA-2, green triangle). The scale bars are 2 nm for a and 0.5 nm for B.

We then tested the HER performance of a Mo SAs/ML-MoS$_2$ catalyst which lay on a SiO$_2$/Si substrate. Unlike conventional powder catalysts, the key feature of the Mo SAs/ML-MoS$_2$ catalyst is that we can accurately determine the areas and numbers of active sites by using a microreactor design (**Figure 3A**). A typical two-terminal device was fabricated using e-beam lithography, where a pristine ML-MoS$_2$ flake were covered by polymethyl methacrylate (PMMA) with exposure of certain areas of the basal plane and two electrode pads (see device fabrication details in Methods and **Figure S18**).[15-17] The PMMA layer not only protects the device from acidic or alkaline electrolytes (**Figures S19 and S20**), but it also ensures that only the pre-defined exposed area of catalyst is in contact with the electrolyte. After the hydrogen plasma treatment, only the exposed MoS$_2$ area has been converted into Mo SAs/ML-MoS$_2$ (**Figure 3B and Figure S21**). A microreactor was used for electrocatalytic HER measurements of the MoS$_2$ flake exposed to both acidic (0.5 M H$_2$SO$_4$) and alkaline (1.0 M KOH) medium (**Figures 3C and 3D**). Notably, in this experiment, Mo SA/ML-MoS$_2$ came from the pristine

MoS$_2$ sample, which ensures the same surface area and device design. Therefore, we were able to directly study how the new Mo SA sites in Mo SAs/ML-MoS$_2$ contributed to the HER.

We first tested Pt-based electrocatalysts in the same microreactor design, which showed comparable results to a normal electrochemical reactor, suggesting the good reliability of the microreactor (**Figure S19**). For MoS$_2$-based catalysts, the polarization curves of the pristine MoS$_2$ and Mo SAs/ML-MoS$_2$ samples in both H$_2$SO$_4$ and KOH solutions are shown in **Figure 3E**. The catalytic performance of the pristine MoS$_2$ was comparable to that of monolayer MoS$_2$ catalysts recently reported (**Table S1**). Impressively, we found that the Mo SAs/ML-MoS$_2$ sample requires much smaller overpotentials to reach the same current densities than the pristine MoS$_2$. For example, the basal plane of Mo SAs/ML-MoS$_2$ shows remarkable HER performance in acidic medium, with an overpotential of 107 mV (versus RHE) at a current density of 10 mA cm$^{-2}$ and an overpotential of 261 mV (versus RHE) at a current density of 400 mA cm$^{-2}$, which are among the best values in 2D MoS$_2$-based electrocatalysts (**Table S1**). At overpotentials of 150 mV or 200 mV, the current densities of the Mo SAs/ML-MoS$_2$ are 30-times or 180-times higher than that of the pristine MoS$_2$, respectively. In alkaline medium, the Mo SAs/ML-MoS$_2$ sample also showed decent HER performance with an overpotential of 209 mV at a current density of 10 mA cm$^{-2}$, which is smaller than the pristine MoS$_2$ sample (322 mV at 10 mA cm$^{-2}$). We note that MoS$_2$ electrocatalysts show a slower charge transfer process in alkaline than in acidic medium, presumably related to sluggish water dissociation on the MoS$_2$ surface in alkaline medium, as previously reported.[9,43,44] The overpotentials required to achieve a current density of 10 mA cm$^{-2}$ for different samples are summarized in **Figure S22A.** The device remains good after all the tests, suggesting that only the exposed basal plane of MoS$_2$ was involved in the electrochemical process (**Figure S23**).

The model electrocatalysts using the ML-MoS$_2$ on the substrate provide unique platforms for studying the HER kinetics of Mo SA sites. It is known that the kinetics analyses of heterogeneous reactions require an accurate interfacial area, however this is challenging to quantify in practical situations. For conventional electrochemical cells, catalysts are loaded on conductive substrates in powder or film forms, resulting in a ratio of interfacial surface area to surface area of electrode ($A_{int}/A_{ele}$) that is larger than one, but difficult to determine accurately. Direct use of geometrical surface area will overestimate the activity of a catalyst.[45] Other methods such as electrochemical surface area (ECSA) measurements have limitations for complex catalysts and can be affected by testing conditions. Here, we combined the microreactor design and monolayer nature of the MoS$_2$ on a substrate with encapsulation to make $A_{interface}/A_{electrode}$ close to one, making it feasible to quantitatively study the HER kinetics of Mo SAs/ML-MoS$_2$ (**Figure S1**). To analyze the rate-determining steps of catalysts, the Butler-Volmer equation and its derivations are usually used. Using the Tafel plots of the two samples, the Tafel slopes were obtained from the linear region (**Figure 3F**). The pristine MoS$_2$ sample shows Tafel slopes of 79.2 mV dec$^{-1}$ in acid and 108.4 mV dec$^{-1}$ in alkali, similar to other reports.[15,33,35] Based on previous studies,[8,9,34,46] the rate-determining step for the pristine MoS$_2$ catalysts should be the Volmer step,[9] a step of the primary discharge of protons. In sharp contrast, the Mo SAs/ML-MoS$_2$ catalyst shows very small Tafel slops of 35.1 mV dec$^{-1}$ in acid and 36.4 mV dec$^{-1}$ in alkali, close to the theoretical value of 40 mV dec$^{-1}$ at which the Heyrovsky reaction pathway is the rate-determining step. The lower Tafel slope of Mo SAs/ML-MoS$_2$ than pristine MoS$_2$ indicates fast HER kinetics for the Mo SAs/ML-MoS$_2$. Tafel slopes for different samples are summarized in **Figure S22B**. The fast HER kinetics of the Mo SAs/ML-MoS$_2$ is further confirmed by the TOF analyses (**Note S1 and Figure S22C**). For example, the TOF of the Mo SAs/ML-MoS$_2$ catalyst is as high as ~10$^3$ s$^{-1}$ at 100 mV. In addition, the samples do not show obvious changes after repeated tests (**Figure S24**). These results indicate that Mo SAs/ML-MoS$_2$ is a new type of active site for MoS$_2$ that shows fast HER kinetics.

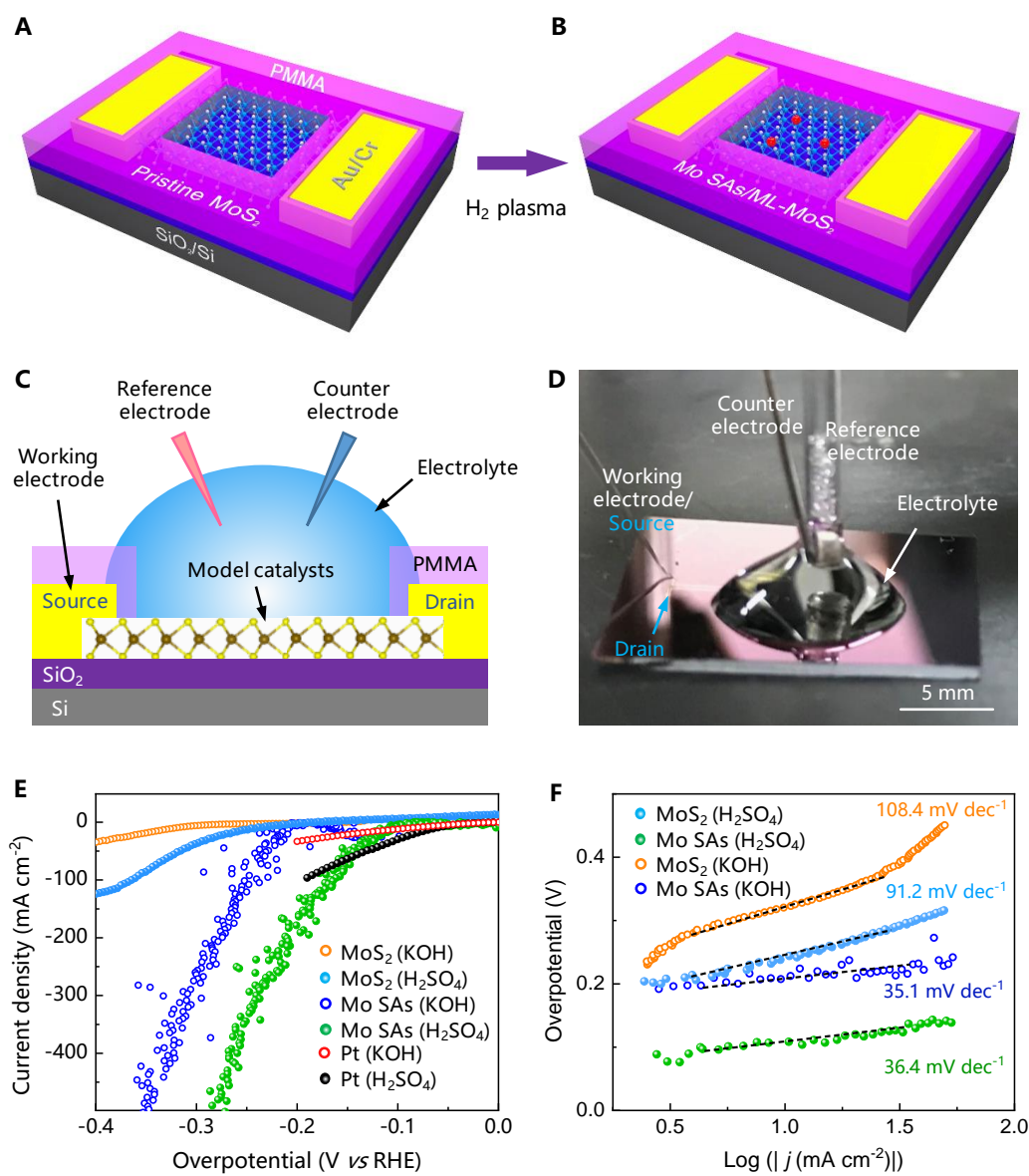

**Figure 3. Electrochemical performance of the Mo SAs/ML-MoS$_2$ catalysts.** (A,B) Schematics showing a ML-MoS$_2$ flake (A) before and (B) after hydrogen plasma treatment, where only a certain area of the MoS$_2$ was exposed for hydrogen plasma treatment and subsequent electrochemical tests. (C) Schematic and (D) optical image showing the electrochemical microreactor based on a ML-MoS$_2$ flake. (E) Polarization curves and (F) Tafel plots of the pristine ML-MoS$_2$, Mo SAs/ML-MoS$_2$, and Pt

samples in $H_2SO_4$ (0.5 M) and KOH (1.0 M). Interfacial surface areas used for calculation of current densities were obtained from the exposed surface area of each device, which was accurately determined during the electron beam lithography process.

**DISCUSSION**

We carried out DFT calculations to elucidate the effects of unsaturated Mo SAs on the catalytic activity of Mo SAs/ML-$MoS_2$. Mo atoms in intact $MoS_2$ monolayer were also included for a better understanding of this issue. The relaxed structures of the pristine $MoS_2$, Mo SA-1, S vacancy, and Mo SA-2 are shown in **Figures 4A-D**. Here, the calculated Gibbs free energies of hydrogen adsorption ($\Delta G_H$) on these structures are shown in **Figure 4E**, with their optimized configurations shown in **Figure S25**. It is generally accepted that an almost zero $\Delta G_H$ shows a good HER electrocatalyst.[47] A large positive/negative $\Delta G_H$ suggests the existence of a large energy barrier in the processes of hydrogen adsorption/desorption. We found that hydrogen is difficult to absorb on Mo sites in a pristine $MoS_2$ basal plane with a large $\Delta G_H$ of 2.356 eV, leading to sluggish HER kinetics (**Figure 4E**). In contrast, Mo SAs on $MoS_2$ are found to be efficient sites for HER, because hydrogen can easily absorb and desorb with small hydrogen free energy of 0.003 eV (Mo SA-1 site) and −0.114 eV (at Mo SA-2 site), which are comparable with or even smaller than 0.041 eV for $MoS_2$ with a S vacancy, a well-recognized HER active site.[25] These results suggest that isolated Mo SAs on a $MoS_2$ surface, especially the Mo SA-1, is a new active center for ultrafast hydrogen evolution.

To further understand the improved catalytic activity of Mo SAs/ML-$MoS_2$ compared to the pristine $MoS_2$, we analyzed the projected density of states (PDOS) of adsorbed hydrogen atoms on the different structures shown in **Figures 4A-D**. For the pristine $MoS_2$, the PDOS of H shows well-isolated sharp peaks below the Fermi level, indicating weak bonding behavior of H. Accordingly, the Fermi level is pinned at the conduction band minimum of MoS$_2$, which is consistent with an earlier report that considered the lowest unoccupied state as the indicator for the bonding strength of H.[48] In contrast, for the other three cases, there are clear hybridized bonding and anti-bonding states below and above the Fermi level with wider peaks. These different bonding characterizations are easily seen from the typical partial charge density distributions (insets of **Figure 4F**), for the hybridized bonding states of a hydrogen atom adsorbed in the different cases. Hybridization between H and Mo orbitals not only increases chemical bonding but lowers the Fermi level compared to the pristine case (indicated by vertical arrows in **Figure 4F**), both of which lead to a lower $\Delta G_H$.

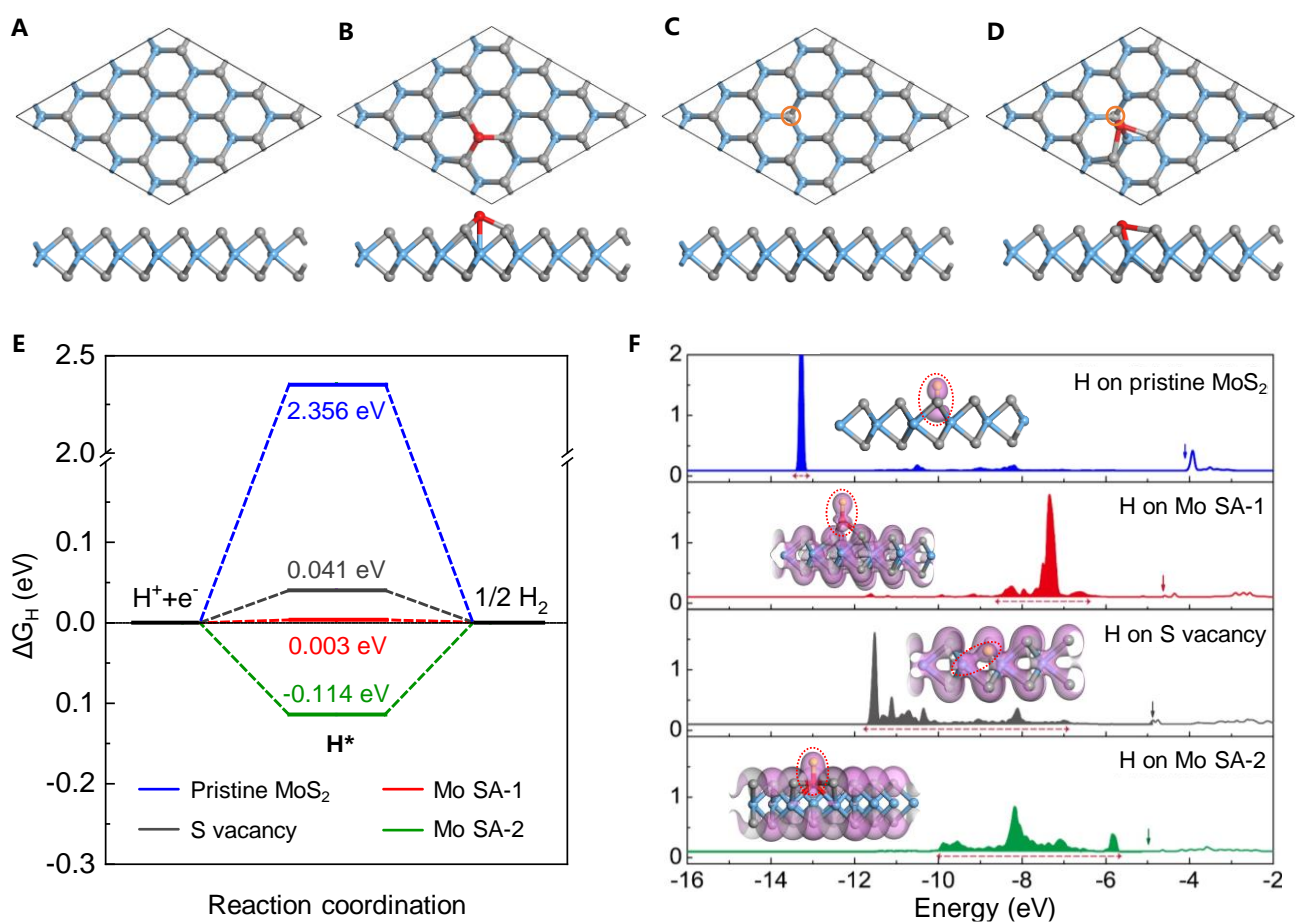

**Figure 4. Theoretical analyses.** (A-D) Top view and side view of the slab models used to describe the

(A) pristine MoS$_2$, (B) Mo SA-1, (C) S vacancy, and (D) Mo SA-2. Atoms in grey and blue represent the S and Mo atoms of the MoS$_2$, Mo SAs are highlighted in red, and S vacancies are highlighted by orange circles. e Free energy profiles of HER at the equilibrium potential for Mo atoms at the basal plane of the pristine MoS$_2$ (blue line), S vacancy (grey line), Mo SA-1 (red line), and Mo SA-2 (green line). Compared to saturated Mo atoms in the basal plane, S vacancies and Mo SA/ML-MoS$_2$ are unsaturated and show fast HER kinetics theoretically. (F) PDOS of adsorbed hydrogen atoms for different MoS$_2$ configurations, *i.e.*, pristine (blue line), S vacancy (grey line), Mo SA-1 (red line) and Mo SA-2 (green line). The occupied states are filled with different colors, and unoccupied states are unfilled. The vertical arrows indicate the Fermi level. The energy levels are referenced to the vacuum level. The insets from top to bottom in panel F are the partial charge density distribution for characteristic bonding states in the energy range that indicated by horizontal dashed arrows for different cases. The isovalues of charge density are 0.1 e/Å$^3$ for pristine MoS$_2$ and the Mo SA-1 cases and 0.25 e/Å$^3$ for the S vacancy and Mo SA-2 cases to clearly display the bonding states between hydrogen atoms and substrates, which are highlighted by red dotted ellipses.

In conclusion, we developed a cold hydrogen plasma reduction method and syntehsized different metal single atoms on 2D monolayer systems. We further fabricated a 2D model electrocatalyst made of Mo single atoms on a cogenetic monolayer MoS$_2$, which shows a superior improvement of the HER activity compared to pristine MoS$_2$. Experimental and theoretical results indicate that Mo single atoms on monolayer MoS$_2$ are a new type of active site for HER with ultrafast hydrogen adsorption/desorption kinetics, originating from a strong hybridization between the single Mo atoms and hydrogen atoms. This work opens an alternative strategy to design high performance MoS$_2$-based catalysts to generate hydrogen by electrochemical water splitting. The cold plasma reduction method addressed the current challenge of synthesizing high loading amounts of single atoms on 2D monolayers. In addition, other metal single atoms could in principle be synthesized from their corresponding transition metal chalcogenides to make a family of metal single atoms/2D monolayers, since they show similar formation free energies of their metal atoms with $MoS_2$ and $WS_2$.[40] The formation of single atoms on monolayer 2D materials with clear and well-defined structures offers valuable platforms to study various electrochemical processes beyond hydrogen evolution discussed here, including lithium-sulphur batteries, fuel cells, and photoelectrochemical devices.

**EXPERIMENTAL PROCEDURES**

Full details of experimental procedures are provided in the Supplemental Information.

**SUPPLEMENTAL INFORMATION**

Supplemental Information includes Supplemental Experimental Procedures, 25 figures, and 3 tables and 1 note can be found with this article online.

**AUTHOR CONTRIBUTIONS**

Y.L. and B.L. conceived the idea and directed the research. M.L., Q.F., and Z.C. synthesized $MoS_2$ flakes by CVD method. L.T. synthesized $WS_2$ flakes by CVD method. Y.L. and S.X. performed the plasma treatment experiments under guidance of W. L.. Y.L. and L.T. performed Raman, SEM, XPS, and AFM characterization. H.P. and Y.L. fabricated microcell devices under guidance of Y.Z.. B.D. and U.K. helped in electronic device measurements. Y.L. performed the electrochemical tests with help from B.D.. Z.G. and J.L. performed TEM experiments. S.Z. and X.Z. performed the theoretical calculations. Y.L, S.Z., J.L., X.Z., H.M.C., and B.L. analyzed the data, interpreted the results and wrote the manuscript with feedbacks from the other authors.

**ACKNOWLEDGMENTS**

We sincerely thank Prof. Andre Geim for fruitful discussions and inputs to this work. We also thank Prof. Hengqiang Ye for the analyses of TEM data, and Chi Zhang, Zhiyuan Zhang, and Dr. Qiangmin Yu for help in the electrochemical measurements. We acknowledge support by the National Natural Science Foundation of China (Nos. 51722206 and 11674150), the Youth 1000-Talent Program of China, Guangdong Innovative and Entrepreneurial Research Team Program (Nos. 2017ZT07C341 and 2016ZT06D348), the Economic, Trade and Information Commission of Shenzhen Municipality for the "2017 Graphene Manufacturing Innovation Center Project" (No. 201901171523), Shenzhen Science and Technology Innovation Commission (JCYJ20160613160524999) and the Development and Reform Commission of Shenzhen Municipality for the development of the "Low-Dimensional Materials and Devices" discipline.